\documentclass[aps,pra,reprint,showpacs,superscriptaddress,longbibliography]{revtex4-1}
\usepackage{amsmath,amssymb,graphicx,units}
\usepackage[usenames,dvipsnames]{color}
\usepackage[plainpages=false,pdfpagelabels,colorlinks=true,linkcolor=red,
urlcolor=blue,citecolor=blue,pdftitle={Title},pdfauthor={},pdfdisplaydoctitle=true,
pdfduplex=DuplexFlipLongEdge]{hyperref}
\usepackage{CJKulem}

\newcommand{\abs}[1]{\ensuremath{\left| #1 \right|}}
\newcommand{\ket}[1]{\left\vert#1\right\rangle}

\newcommand{\braket}[2]{\left\langle#1\left\vert#2\right.\right\rangle}

\newcommand{\vev}[1]{\left\langle #1\right\rangle}
\newcommand{\1}{\mbox{\bf 1}}
\makeatother
\begin{document}

\title{Phase transitions in the Haldane-Hubbard model with ionic potentials}
\author{Hao Yuan}
\affiliation{Institute of Ultrafast Optical Physics, Department of Applied Physics $\&$ MIIT Key Laboratory of Semiconductor Microstructure and Quantum Sensing, Nanjing University of Science and Technology, Nanjing 210094, China}

\author{Yangbin Guo}
\affiliation{Institute of Ultrafast Optical Physics, Department of Applied Physics $\&$ MIIT Key Laboratory of Semiconductor Microstructure and Quantum Sensing, Nanjing University of Science and Technology, Nanjing 210094, China}

\author{Ruifeng Lu}
\affiliation{Institute of Ultrafast Optical Physics, Department of Applied Physics $\&$ MIIT Key Laboratory of Semiconductor Microstructure and Quantum Sensing, Nanjing University of Science and Technology, Nanjing 210094, China}

\author{Hantao Lu}
%\email{luht@lzu.edu.cn}
\affiliation{School of Physical Science and Technology $\&$ Lanzhou Center for Theoretical Physics, Key Laboratory of Theoretical Physics of Gansu Province, Lanzhou University, Lanzhou, 730000, China}

\author{Can Shao}
\email{shaocan@njust.edu.cn}
\affiliation{Institute of Ultrafast Optical Physics, Department of Applied Physics $\&$ MIIT Key Laboratory of Semiconductor Microstructure and Quantum Sensing, Nanjing University of Science and Technology, Nanjing 210094, China}

% \date{\today}

\begin{abstract}
By employing the exact-diagonalization method, we revisit the ground-state phase diagram of the Haldane-Hubbard model on the honeycomb lattice with staggered sublattice potentials. The phase diagram includes the band insulator, Mott insulator, and two Chern insulator phases with Chern numbers $C=2$ and $C=1$, respectively. The character of transitions between different phases is studied by analyzing the lower-lying energy levels, excitation gaps, structure factors, and fidelity metric. We find that the $C=1$ phase can be continuously deformed into the $C=2$ phase without a gap closure in the periodic boundary condition, while a further analysis on the Berry curvatures indicates that the excitation gap closes at the phase boundary in a twisted boundary condition, accompanied by the discontinuities of structure factors. All the other phase transitions are found to be first-order ones as expected.

\end{abstract}

\maketitle

\section{Introduction}\label{sec:intro}

As one of the most important concepts in condensed matter physics, the spontaneous symmetry breaking plays a key role in the Ginzburg-Landau theory for describing the phase transitions of matter~\cite{LANDAU1980446}. While in the 1980s, the discovery of the integer quantum Hall effect in a two-dimensional electron gas under strong magnetic field provided another paradigm to identify quantum states with topological characters~\cite{PhysRevLett.45.494}.
%Subsequently, more exotic states such as quantum anomalous Hall effect~\cite{Haldane1988} and quantum spin Hall effect~\cite{Kane05, Bernevig06} have attracted much attention.
In the past few decades, the topological perspective has played an increasingly important role as a new standard for the classification of a large class of materials~\cite{Zhang19, Vergniory19, Tang19, Kruthoff2017}, and the topological ordered states in non-interaction systems have been completely classified based on the time-reversal, particle-hole, chiral, and crystal symmetries~\cite{Schnyder08, Kitaev2009, Budich2013, slager2013}.

In the interacting systems, the correlation effect of electrons is expected to give rise to more intriguing and richer phenomena~\cite{Rachel_2018, Hohenadler_2013}. A famous example is the fractional quantum Hall state, which allows the emergence of quasiparticle excitations that carry fractional charge and obey fractional statistics largely due to the strong repulsion among electrons~\cite{Laughlin83, Stormer99, Nayak08}. Another example is the so-called topological Mott insulator, where the electronic interactions may generate dynamically the spin-orbit coupling, promoting the system into a topological nontrivial state~\cite{Raghu2008, Wen2010, Budich2012, Dauphin2012, Weeks2010, Ruegg2011, Yang2011, Yoshida2014, wang2012}. Despite the fact that some of these mean-field results remain controversial~\cite{Garcia_Martinez13, Daghofer14, Motruk15, Capponi15, Scherer15}, interaction-induced topological states may still be observed in two-dimensional systems with quadratic band crossing and weak interactions~\cite{Sun09, Murray14, Venderbos16, Wu16, Zhu16, Vafek10, Wang17, Uebelacker2011}. The interplay between topology and local orders has been largely addressed by introducing interaction into topological systems, and a number of novel phases have been identified~\cite{Varney11, Shao2021, Vanhala2016, Wang2019, Tupitsyn19, Mertz19, Hetenyi20, Ebrahimkhas21, Ebrahimkhas_2022}. For example, an exotic phase was found recently with one of the spin components in the Hall state and the other one in a localized state, when the inversion-breaking ionic potential is incorporated into the spinful Haldane-Hubbard model~\cite{Vanhala2016, Tupitsyn19, Mertz19, Wang2019}.

In this paper, we adopt the exact-diagonalization (ED) method to revisit the spinful Haldane-Hubbard model at half filling and zero temperature. By tuning the strength parameters of the staggered sublattice potential $\Delta$ and the onsite Coulomb interaction $U$, we explore its ground-state phase diagram. The resulting phase diagram, including the band insulator, the Mott insulator, and the topological insulators with the Chern numbers $C=1,2$, largely coincides with the ED result obtained by Vanhala \textit{et al.}~\cite{Vanhala2016}. However, for the location of the intermediate exotic phase with $C=1$, due to different clusters used in the calculation, our results show slight differences compared to the ones in Ref.~\cite{Vanhala2016}. Meanwhile, the nature of phase transitions is examined by analyzing the excitation gaps, structure factors and fidelity metrics obtained by the ED method, which provides unbiased information of the finite system we adress. In general, the transition between distinct topological phases should involve a bulk band gap closing in order to accommodate the variation of topological numbers~\cite{Hasan2010,Ranjith_2022,Varney10,Shao2021}. In our case, a seemingly continuous phase transition without gap closing appears between the two topological nontrivial phases ($C=1$ and $C=2$) in the periodic boundary condition (PBC), even though the cluster we chose has already contained the $K$ high-symmetry point in its reciprocal lattice~\cite{Varney10,Shao2021}. A careful examination on the singularity of the Berry curvature with finer discretization indicates that the gap indeed does close in a twisted boundary condition (TBC). Under this TBC, the characteristic features of first-order phase transition are restored. Apart from that, all the other phase transitions are found to be first-order ones under PBC as usual.

The presentation is organized as follows: In Sec.~\ref{sec:model}, we introduce the model and all of the quantities we use to characterize the different phases. Section \ref{sec:ED_results} presents the results of the exact-diagonalization method. Lastly, the conclusion is drawn in Sec.~\ref{sec:conclusion}.

\section{Model and measurements} \label{sec:model}
We write the Hamiltonian of the Haldane-Hubbard model as $\hat H=\hat H_k+\hat H_l$, where the kinetic part is

\begin{eqnarray}
\hat H_k=&-&t_1\sum_{\langle i,j\rangle,\sigma}(\hat c^{\dagger}_{i,\sigma} \hat c^{\phantom{}}_{j,\sigma}+\text{H.c.}) \nonumber \\
&-& t_2\sum_{\langle\langle i,j\rangle\rangle,\sigma}(e^{{\rm i}\phi_{ij}}\hat c^{\dagger}_{i,\sigma} \hat c^{\phantom{}}_{j,\sigma}+\text{H.c.})
\label{eq:H1}
\end{eqnarray}
and the local part is
\begin{eqnarray}
\hat H_l=&-&U\sum_{i}\hat n_{i,\uparrow}\hat n_{i,\downarrow}+\Delta\sum_{i,\sigma}{\rm sgn}(i)\hat n_{i,\sigma}.
\label{eq:H2}
\end{eqnarray}
Here, $\hat c^{\dagger}_{i,\sigma}$ ($\hat c^{\phantom{}}_{i,\sigma}$) creates (annihilates) an electron at site $i$ of the honeycomb lattice with spin $\sigma=\uparrow {\rm or} \downarrow$, and $\hat n_{i,\sigma}\equiv \hat c_{i,\sigma}^\dagger\hat c_{i,\sigma}^{\phantom{}}$ is the number operator. $t_1$ ($t_2$) is the nearest-neighbor (next-nearest-neighbor) hopping constant. In the local interaction part, $U$ is the on-site interaction strength, and $\Delta$ is the magnitude of the staggered potential with $\mathrm{sgn}(i)$ being $+1$ for sublattice A and $-1$ for sublattice B, respectively. The phase $\phi_{i,j}=\phi$ ($-\phi$) in the clockwise (anticlockwise) loop is introduced to the second hopping term. Throughout the paper, we focus on the ground-state phase diagram at half filling.

Several properties can be used to characterize the quantum phase transition. One of them is the ground-state fidelity metric, which is defined as~\cite{Zanardi06,CamposVenuti07,Zanardi07} :
\begin{eqnarray}
g(U,\delta U)\equiv\frac{2}{N }\frac{1-\abs{\braket{\Psi_0(U)}{\Psi_0(U+\delta U)}}}{(\delta U)^2},
\label{eq:g}
\end{eqnarray}
with $\ket{\psi_0(U)}[\ket{\psi_0(U+\delta U)}]$ being the ground state of $\hat H(U)[\hat H(U+\delta U)]$. In what follows, we set $\delta U = 10^{-3}$. For first-order transitions in finite-size systems, this quantity is expected to exhibit a distinguished and sharp peak or a discontinuity at the critical point; while for continuous transitions there will be instead a wider and smaller "hump"~\cite{Varney10,Imriska16,Shao2021}.

To characterize the Mott insulator (MI) and band insulator (BI) respectively, the spin-density wave (SDW) and charge-density wave (CDW) structure factors can be applied. We define them in a staggered fashion as follows:
\begin{eqnarray}
S_{\mathrm{SDW}} = \frac{1}{N}\sum\limits_{i,j}{(-1)}^{\eta}  \vev{(\hat n_{i,\uparrow}-\hat n_{i,\downarrow}) (\hat n_{j,\uparrow}-\hat n_{j,\downarrow})}, \nonumber \\
S_{\mathrm{CDW}} = \frac{1}{N}\sum\limits_{i,j}{(-1)}^{\eta}  \vev{(\hat n_{i,\uparrow}+\hat n_{i,\downarrow})(\hat n_{j,\uparrow}+\hat n_{j,\downarrow})},
\label{eq:S}
\end{eqnarray}
where $\eta = 0$ ($\eta = 1$) indicates that sites $i$ and $j$ are in the same (different) sublattice.

The topological invariant is the Chern number. It can be evaluated by integration over the TBCs~\cite{Niu85,Didier91}:
\begin{align}
  C = \int \frac{d\phi_x d\phi_y}{2 \pi {\rm i}} \left( \braket{\partial_{\phi_x}
      \Psi}{\partial_{\phi_y} \Psi}- \braket{\partial_{\phi_y}
      \Psi}{\partial_{\phi_x} \Psi} \right),
\label{eq:C}
\end{align}
where $\ket{\Psi}$ is the ground-state wave function, and $\phi_x$ ($\phi_y$) is the twisted phase along the $x$ ($y$) direction. To avoid the integration with huge computational resources, we instead use a discretized version~\cite{Fukui05, Varney11, Zhang13} in which $\Delta\phi_x=\frac{2\pi}{N_x}$ and $\Delta\phi_y=\frac{2\pi}{N_y}$. The numerical computation is then discretized in $N_x$ and $N_y$ intervals, and the ground state can be specified as $|\Psi^0_{m,n}\rangle$ with $m\in[0,N_x)$ and $n\in[0,N_y)$. The discrete version of Berry curvature reads
% \begin{align}
%   F_{m,n} = -i\ \rm{log}\left( \frac{U^{\emph{x}}_{\emph{m,n}}U^{\emph{y}}_{\emph{m}+1,\emph{n}}}{U^{\emph{x}}_{\emph{m,n}+1}U^{\emph{y}}_{\emph{m,n}}} \right),
% \label{eq:F_mn}
% \end{align}
\begin{align}
F_{m,n} = -i\ \rm{log}\left( \frac{U^{{x}}_{{m,n}}U^{{y}}_{{m}+1,{n}}}{U^{{x}}_{{m,n}+1}U^{{y}}_{{m,n}}} \right),
\label{eq:F_mn}
\end{align}
where the complex entries in the $U$ matrices are
% \begin{align}
%   U^{\emph{x}}_{m,n} = \frac{\langle\Psi^{0}_{\emph{m,n}}|\Psi^{0}_{\emph{m}+1,\emph{n}}\rangle}{|\langle\Psi^{0}_{\emph{m,n}}|\Psi^{0}_{\emph{m}+1,\emph{n}}\rangle|}, U^{\emph{y}}_{m,n} = \frac{\langle\Psi^{0}_{\emph{m,n}}|\Psi^{0}_{\emph{m},\emph{n}+1}\rangle}{|\langle\Psi^{0}_{\emph{m,n}}|\Psi^{0}_{\emph{m},\emph{n}+1}\rangle|}.
% \label{eq:U_mn}
% \end{align}
\begin{equation}
U^{{x}}_{m,n} = \frac{\braket{\Psi^{0}_{{m,n}}}{\Psi^{0}_{{m}+1,{n}}}}{\abs{\braket{\Psi^{0}_{{m,n}}}{\Psi^{0}_{{m}+1,{n}}}}}, \quad
U^{{y}}_{m,n} = \frac{\braket{\Psi^{0}_{{m,n}}}{\Psi^{0}_{{m},{n}+1}}}{\abs{\braket{\Psi^{0}_{{m,n}}}{\Psi^{0}_{{m},{n}+1}}}}.
\label{eq:U_mn}
\end{equation}

The Chern number can be written as the summation of the discretized Berry curvatures, i.e., $C = \sum_{m,n}\frac{F_{m,n}}{2\pi}$, with the values of $F_{m,n}$ being chosen in the principal branch $(-\pi,\pi]$. It has been shown to converge to the true Chern number if sufficient twisted phases are used.

In what follows, $t_1$ is set to be the unit of energy and $t_2=0.2$. We further fix the Haldane phase $\phi$ to $\pi/2$ in order to maximize the Chern insulator (CI) phase~\cite{Haldane1988, Varney10}.

\section{Results and analysis}\label{sec:ED_results}

By employing the Arnoldi~\cite{Lehoucq97arpack} method, we can obtain the ground state and several low-lying excited states of the Haldane-Hubbard model. The $12A$ cluster we use can be found in Ref.~\cite{Shao2021}, together with the corresponding $k$ points in the reciprocal space. It has been noticed that compared to other $12$- and $16$-site lattices, this cluster, whose reciprocal lattice contains the $\Gamma$ point, all the $K$ points, and one pair of $M$ points, is more suitable for the purpose of ED analysis in the interacting Haldane model~\cite{Shao2021}. The smallest cluster containing $K$ points other than $12A$ is the $18A$ lattice (though containing no $M$ points), which is not accessible with our current computational resources and can be explored in the future studies.

\begin{figure}[t]
\centering
\includegraphics[width=0.48\textwidth]{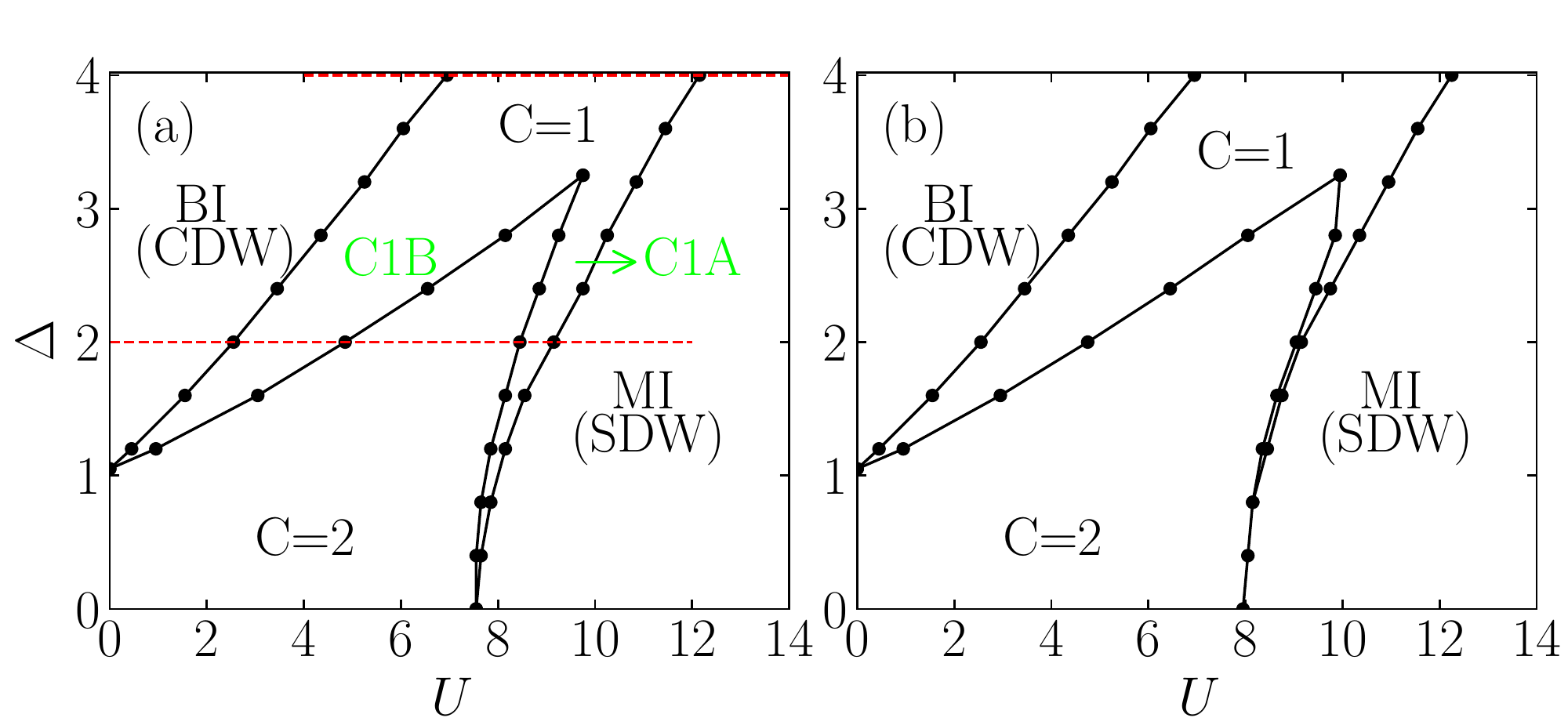}
\caption{Phase diagram in the parametric space ($U$, $\Delta$) of the Haldane-Hubbard model based on results of Chern number, with the meshes of (a) $6\times6$ and (b) $20\times20$ twisted phases. The red dashed lines denote the parameters we choose to show more details below. The lattice of $12A$ cluster is adopted for the ED calculations.
}
\label{fig_1}
\end{figure}

\begin{figure}
\centering
\includegraphics[width=0.48\textwidth]{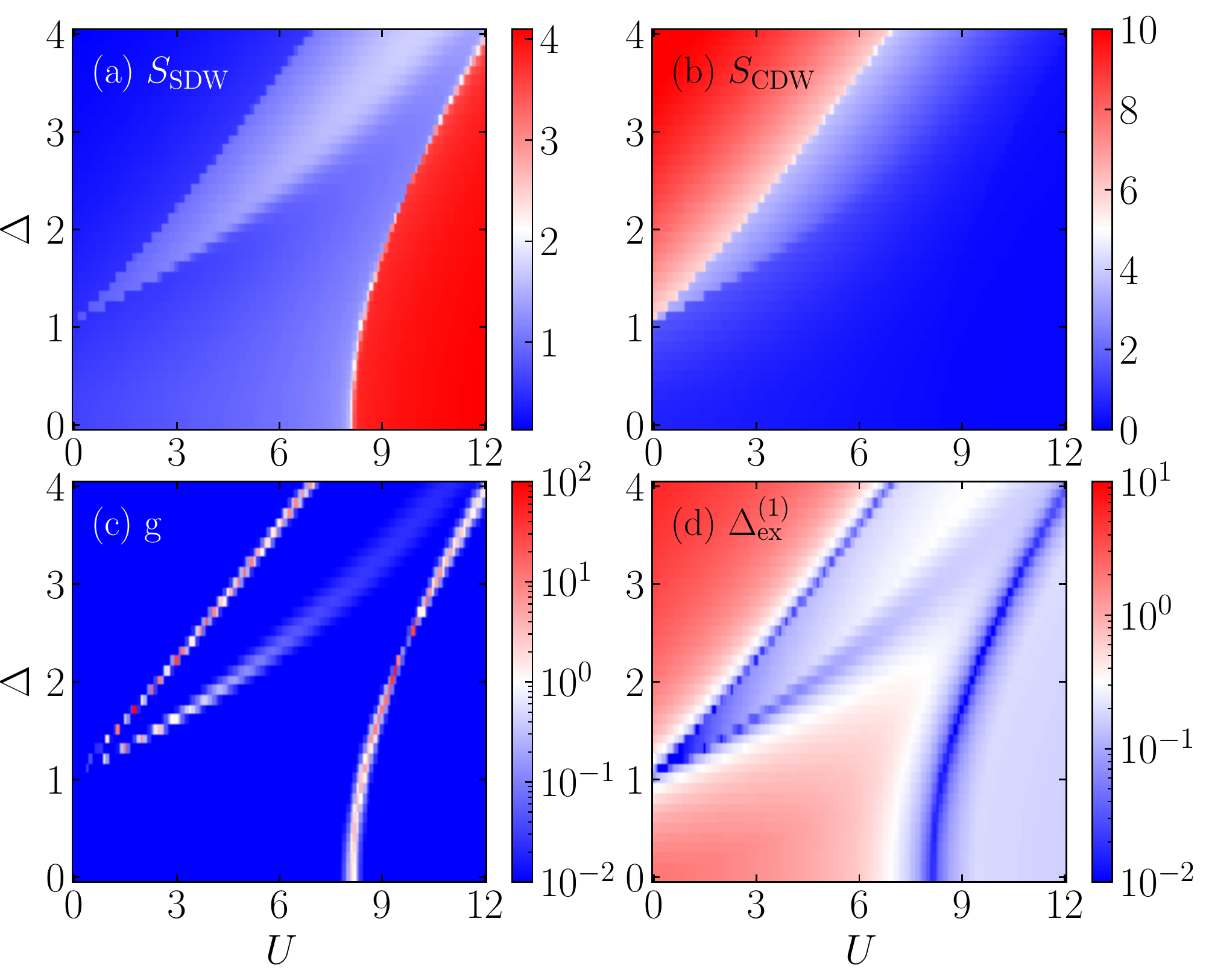}
\caption{Contour plots of (a) the SDW structure factor $S_{\rm SDW}$, (b) the CDW structure factor $S_{\rm CDW}$, (c) the fidelity metric $g$ and (d) the first excitation gap $\Delta_{\mathrm{ex}}^{(1)}$ as a function of $U$ and $\Delta$.
}
\label{fig_2}
\end{figure}

The ($U$, $\Delta$) phase diagrams based on the results of Chern number obtained by $6\times6$ and $20\times20$ meshes in the boundary phase space are shown in Figs.~\ref{fig_1}(a) and \ref{fig_1}(b), respectively. They both agree qualitatively to the phase diagram in Ref.~\cite{Vanhala2016}: the CI phase with $C=2$ in the small $U$ and $\Delta$ region; the BI (CDW) and MI (SDW) phases are governed by large $\Delta$ and $U$, respectively. Sandwiched between the BI and MI phases there is another topological nontrivial phase with $C=1$. However, it can be observed that as the meshes increase from $6\times6$ to $20\times20$, one of the regions of the $C = 1$ phase that lies between the $C = 2$ and MI phases (denoted as C1A) reduces considerably and further vanishes when $\Delta\leq0.8$, as shown in Fig.~\ref{fig_1}(b). While on the other hand, another region of $C=1$ that lies between the $C=2$ and BI phases (denoted as C1B) remains intact. The narrowing C1A in Fig.~\ref{fig_1}(b) is quite similar to the result of the dynamical mean-field theory (DMFT) in Ref.~\cite{Vanhala2016}, while the C1B phase is consistent with their ED result on a 16-site cluster [see Fig. 2 in Ref.~\cite{Vanhala2016}]. The sensitivity of the C1A region to the mesh size implies the critical nature of the $C=1$ phase in this parameter space. Compared with the ED result in Ref.~\cite{Vanhala2016}, we speculate that it suffers from the finite size effect more severely than the C1B, and whether it can survive in the thermodynamic limit remains to be solved. As for the C1B phase, it is worth mentioning that both the DMFT~\cite{Vanhala2016} and the diagrammatic Monte Carlo~\cite{Tupitsyn19} predicted the absence of $C=1$ phase in the small-$U$ region. However from our and other ED results, we suggest that the issue of whether the finite sublattice potential can induce the spontaneous SU(2) symmetry breaking with small on-site interaction $U$ in this model may still be open. In Figs.~\ref{fig_2}(a), \ref{fig_2}(b), \ref{fig_2}(c) and \ref{fig_2}(d), the contour plots of the SDW structure factor $S_{\mathrm{SDW}}$, the CDW structure factor $S_{\mathrm{CDW}}$, the fidelity metric $g$ and the first excitation gap $\Delta_{\mathrm{ex}}^{(1)}$ as a function of $U$ and $\Delta$ are shown, respectively. We see that the boundary between $C=2$ and C1B phases can be clearly read and are consistent with Fig.~\ref{fig_1}, while the boundary between $C=2$ and C1A phases can not be observed. Similar deviations between the results of Chern number and other properties (including the fidelity and the structure factors) have also been noticed in Refs.~\cite{Imriska16, Shao2021, Varney11}, and been attributed to the finite-size effect.

\begin{figure}[t]
\centering
\includegraphics[width=0.48\textwidth]{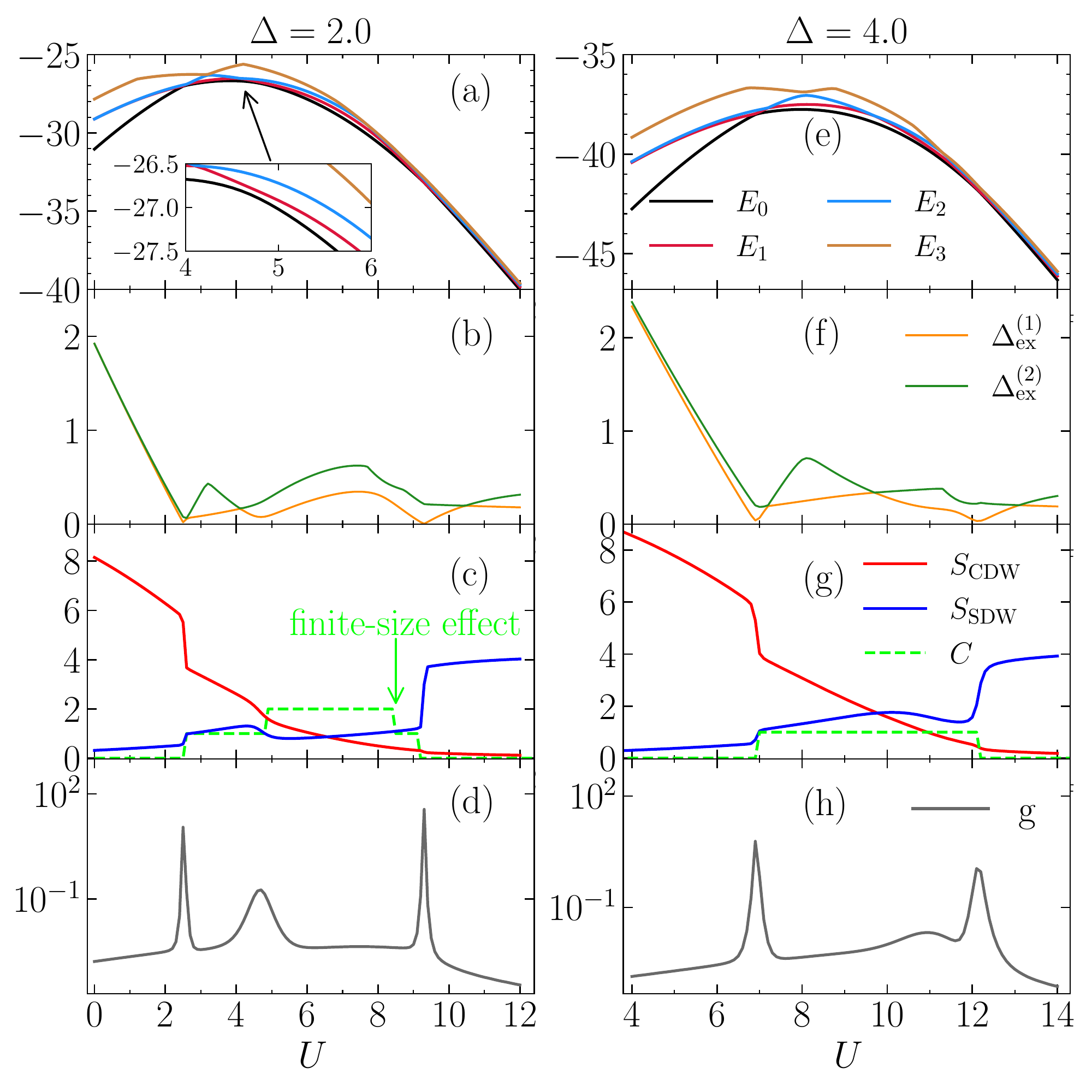}
\caption{(a)(e) Four lowest-lying energy levels $E_{\alpha}$, (b)(f) the excitation gaps $\Delta_{\mathrm{ex}}^{(\alpha)}$ with $\alpha=1,2$, (c)(g) the structure factors $S_{\mathrm{SDW/CDW}}$ and Chern number $C$, and (d)(h) the fidelity metric $g$ of the Haldane-Hubbard model with $\Delta=2.0$ on the left panels and $\Delta=4.0$ on the right panels. The inset in (a) shows a closeup view of the energy levels approaching around $U=4.8$.
}
\label{fig_3}
\end{figure}

In the present study we are mainly interested in the nature of the phase transitions of the model, which has not been fully addressed in the previous works. We choose the ionic potential $\Delta=2.0$ and $\Delta=4.0$ [red dashed lines in Fig.~\ref{fig_1}(a)] to calculate the energy levels $E_{\alpha}$, excitation gaps $\Delta_{\mathrm{ex}}^{(\alpha)}$, structure factors $S_{\mathrm{SDW/CDW}}$ and fidelity metric $g$ as functions of $U$. The results are presented in Fig.~\ref{fig_3}. The first four lowest-lying energy levels are denoted as $E_{\alpha}$ with $\alpha=0, 1, 2, 3$, ($E_0$ is the ground-state energy) and the excitation gap $\Delta_{\mathrm{ex}}^{(\alpha)}$ is defined as $\Delta_{\mathrm{ex}}^{(\alpha)}=E_{\alpha}-E_0$ with $\alpha=1,2,3$. The line of $\Delta=2.0$ crosses successively the BI, C1B, $C=2$, C1A, and MI phases with increasing $U$ from $0.0$ to $12.0$, while the line of $\Delta=4.0$  crosses the BI, $C=2$ and MI phases as $U$ increases from $4.0$ to $14.0$. Note that the green dashed lines in Figs.~\ref{fig_3}(c) and \ref{fig_3}(g) shows the Chern number results of Fig.~\ref{fig_1}(a) with $6\times6$ meshes in the twisted phases. For the phase boundaries of the BI and MI phases ($U=2.55$ and $U=9.15$ for $\Delta=2.0$; $U=6.95$ and $U=12.15$ for $\Delta=4.0$), we can observe the characteristics of first-order phase transition in the vicinity of the critical points, which includes the level crossings between the ground state and one excited state in Figs.~\ref{fig_3}(a) and \ref{fig_3}(e), the vanishing of the first excitation gaps in Figs.~\ref{fig_3}(b) and \ref{fig_3}(f), and the discontinuities of the CDW and SDW structure factors in Figs.~\ref{fig_3}(c) and \ref{fig_3}(g), as well as sharp peaks in the fidelity metrics $g$ in Figs.~\ref{fig_3}(d) and \ref{fig_3}(h). %\shao{These features can be compared with the ED results in Ref.~\cite{Vanhala2016}, where the closure of quasi-particle gap was not observed due to the absence of $K$ point in their $16$-site lattice.}

Nevertheless, some features of a continuous phase transition appear at $(U, \Delta)= (4.75, 2.00)$ which separates the $C=2$ from the C1B phase. Here we leave the transition between $C=2$ and C1A for future investigations, since the region of C1A is much narrower and there are no clear signatures for the transition except for the change of Chern number [see also Fig.~\ref{fig_2}]. As shown by the inset of Fig.~\ref{fig_3}(a), we see that at the transition point, rather than the aforementioned level crossing, the two lowest energy levels just approach each other and then separate with further increasing $U$. Correspondingly, the decreasing of the first excitation gap $\Delta_{\mathrm{ex}}^{(1)}$ takes place instead of gap closing [see Fig.~\ref{fig_3}(b)]. The smooth changing of structure factors instead of finite jump, and a smaller "hump" in fidelity metric instead of sharp peak can be also observed in Figs.~\ref{fig_3}(c) and \ref{fig_3}(d), respectively. As far as we know, in the interacting Haldane systems all phase transitions between topological and locally ordered states are found to be first-order ones if the clusters contains the K high-symmetry point in their reciprocal lattices. \cite{Jotzu2014, Imriska16, Varney11}. For finite-size systems, in general, it is expected that the excitation gap should vanish for some twisted boundary condition if it does not for the usual PBC, in order to accommodate the change of the topological index~\cite{Varney11}. Therefore, the seemingly continuous transition between the $C=2$ and C1B topological phases needs further analysis and following is the result.

\begin{figure}[t]
\centering
\includegraphics[width=0.48\textwidth]{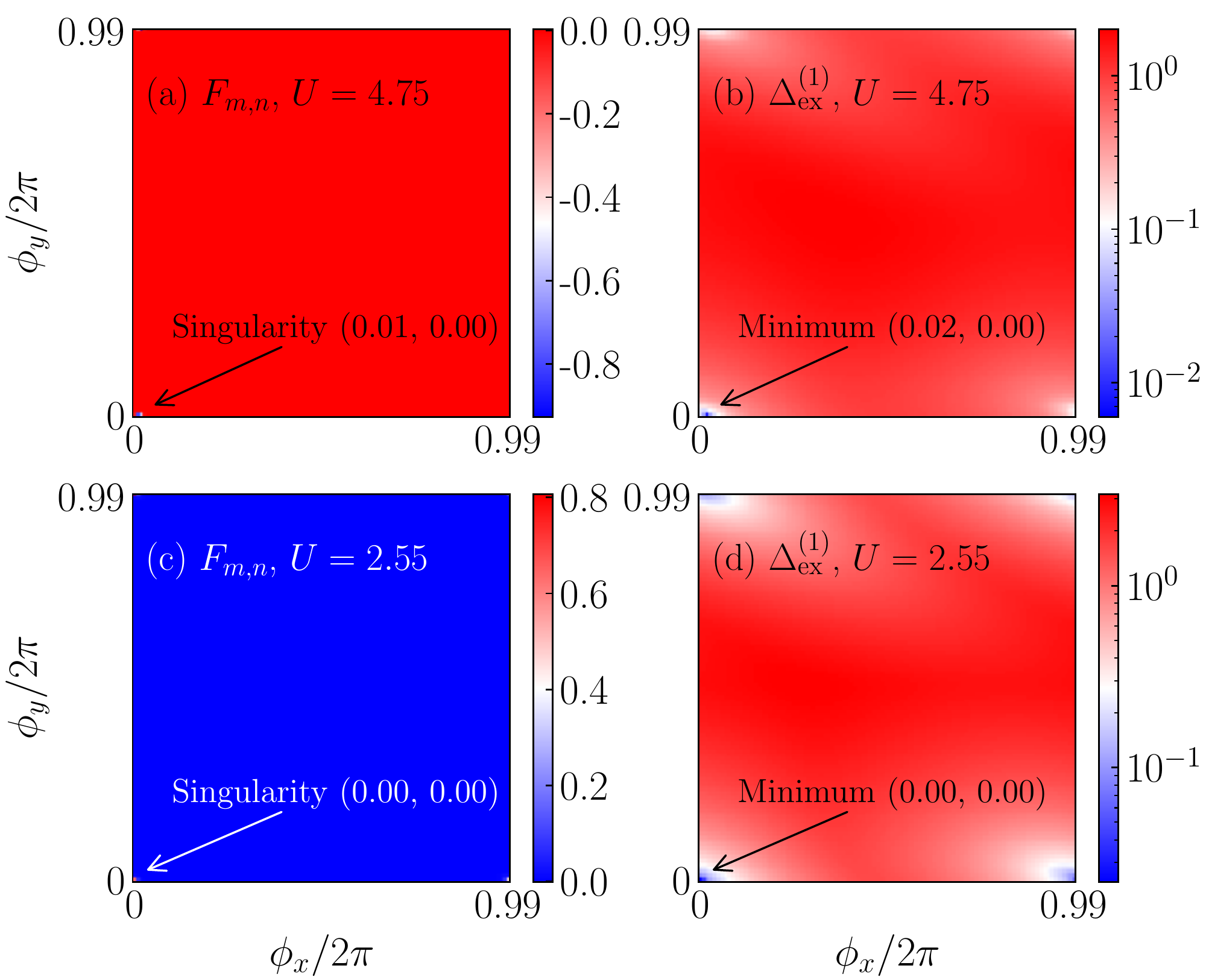}
\caption{(a) The Berry curvature $F_{m,n}$ and (b) the first excitation gap $\Delta_{\rm ex}^{(1)}$ as functions of $(\phi_x/2\pi, \phi_y/2\pi)$ for $(U, \Delta)= (4.75, 2.00)$. (c) $F_{m,n}$ and (d) $\Delta_{\rm ex}^{(1)}$ for $(U, \Delta)= (2.55, 2.00)$.
}
\label{fig_4}
\end{figure}

We use a finer mesh of $100\times100$ in the twisted phase space ($\phi_x$, $\phi_y$) to obtain the discrete Berry curvatures $F_{m,n}$ (defined by Eq.~(\ref{eq:F_mn})) for the critical point $(U, \Delta)= (4.75, 2.00)$. The result is presented in Fig.~\ref{fig_4}(a) as a function of twisted angles, together with the first excitation gap $\Delta_{\mathrm{ex}}^{(1)}$ presented in Fig.~\ref{fig_4}(b). By careful examination, one can observe a singularity of $F_{m,n}$ at ($\phi_x / 2\pi$, $\phi_y / 2\pi$)=($0.01$, $0.00$), and the closure of the first excitation gap that correspondingly takes place at $(\phi_x / 2\pi, \phi_y / 2\pi)=(0.02, 0.00)$. It confirms the speculation in the previous paragraph. For a comparison, we choose another critical point $(U, \Delta) = (2.55, 2.00)$, which lies at the boundary between the $C=1$ topological phase and the BI ordered phase, to perform the same calculation. From Figs.~\ref{fig_4}(c) and \ref{fig_4}(d), it can be observed that both the singularity of $F_{m,n}$ and the minimum of $\Delta_{\mathrm{ex}}^{(1)}$ (gap closing) occur at $(\phi_x / 2\pi, \phi_y / 2\pi)=(0.00, 0.00)$, i.e., at the PBC as expected.

\begin{figure}[t]
\centering
\includegraphics[width=0.48\textwidth]{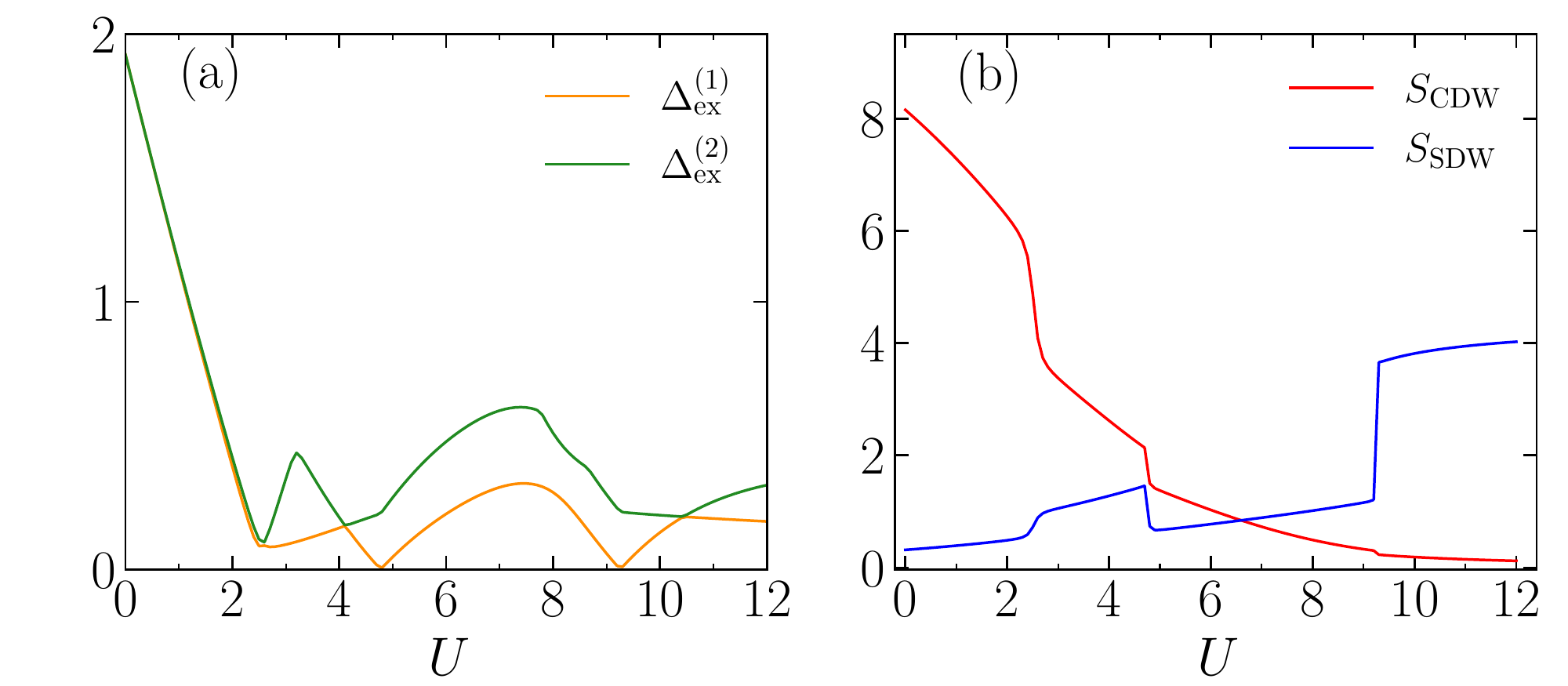}
\caption{(a) The excitation gaps $\Delta_{\mathrm{ex}}^{(\alpha)}$ and (b) the structure factors of SDW and CDW as functions of $U$ in the twisted boundary condition $(\phi_x / 2\pi, \phi_y / 2\pi)=(0.02, 0.00)$ ($\Delta=2.0$).
}
\label{fig_5}
\end{figure}

In order to elucidate the nature of the topological phase transition between the $C=2$ and C1B phases, we calculate under the twisted phase $(\phi_x / 2\pi, \phi_y / 2\pi)=(0.02, 0.00)$ the excitation gaps $\Delta_{\mathrm{ex}}^{(\alpha)}$ and the SDW/CDW structure factors as functions of $U$ with fixed $\Delta=2.0$. The results are shown in Figs.~\ref{fig_5}(a) and (b). We can observe that accompanied with the closure of excitation gap $\Delta_{\mathrm{ex}}^{(1)}$ at $U=4.75$, the discontinuity of structure factors can be recognized. These features indicates that the transition between the topological phases should be also a first-order one.

\section{Summary and Discussion}\label{sec:conclusion}

To summarize, we studied the spinful Haldane-Hubbard model at half-filling on the honeycomb lattice by the ED method. By tuning the on-site interaction strength $U$ and staggered sublattice potential $\Delta$, the ground state exhibits a rich phase diagram, which includes the BI (CDW) and MI (SDW) phases with local order parameters, as well as two topologically nontrivial phases with Chern number $C=2$ and $C=1$. Especially, we observed the energy-level crossings, the discontinuities of structure factors and the sharp peaks of fidelity metric at both BI and MI phase boundaries, signaling first-order phase transitions. Although a nonvanishing of the first excitation gap can be observed for the phase transition between the two topological phases in the periodic boundary condition, the fact that the gap indeed closes in one of the twisted boundary conditions, together with the discontinuity of the structure factors, indicates that the phase transition is also first order.

\begin{acknowledgments}
We would like to thank the anonymous referees for suggesting a careful investigation on the Berry curvature and other valuable comments.
C. S. acknowledges support from the National Natural Science Foundation of China (NSFC; Grant No. 12104229) and the Fundamental Research Funds for the Central Universities (Grant No. 30922010803).
R. F. acknowledges supports from NSFC (Grants No. 11974185) and the Natural Science Foundation of Jiangsu Province (Grant No. BK20170032).
H. L. acknowledges support from NSFC (Grants No. 11874187, No. 12174168 and No. 12047501).
\end{acknowledgments}

\bibliographystyle{apsrev4-1}
%\bibliography{lt}

%merlin.mbs apsrev4-1.bst 2010-07-25 4.21a (PWD, AO, DPC) hacked
%Control: key (0)
%Control: author (72) initials jnrlst
%Control: editor formatted (1) identically to author
%Control: production of article title (-1) disabled
%Control: page (0) single
%Control: year (1) truncated
%Control: production of eprint (0) enabled
%

\end{document}